\documentclass[journal]{IEEEtran}
\usepackage{amsmath}

\usepackage{graphicx}
\usepackage{epstopdf}
\usepackage{amssymb}
\usepackage{subfigure}
\usepackage{cite}
\usepackage{algpseudocode}
\usepackage{algorithm}
\usepackage{multirow}
\usepackage{booktabs}
\usepackage{color}
\usepackage{gensymb}

\ifCLASSINFOpdf
\else
\fi
\begin{document}
%
\title{A closed-loop all-electronic pixel-wise adaptive imaging system for high dynamic range video}
%
%
%

\author{Jie (Jack)~Zhang,
        Jonathan P.~Newman,
        Xiao~Wang,
        Chetan~Singh~Thakur,
		John~Rattray,
		Ralph~Etienne-Cummings,
        and~Matthew~A.~Wilson

\thanks{J. Zhang, J. P. Newman and M. A. Wilson are with the Picower Institute for Learning and Memory, Department of Brain and Cognitive Sciences, Massachusetts Institute of Technology, Cambridge, MA, USA}
\thanks{J. Rattray and R. Etinne-Cummings are with Department
	of Electrical and Computer Engineering, The Johns Hopkins University, Baltimore,
	MD, USA}
\thanks{X. Wang is with Department of Computer Science, Boston University, Boston, MA, USA}
\thanks{C. S. Thakur is with Department of Electronic Systems Engineering, Indian Institute of Science, Bangalore, India}
}
\maketitle

\begin{abstract}

We demonstrated a CMOS imaging system that adapts each pixel’s exposure and sampling rate to capture high dynamic range (HDR) videos.  The system consist of a custom designed image sensor with pixel-wise exposure configurability and a real-time pixel exposure controller. These parts operate in a closed-loop to sample, detect and optimize each pixel’s exposure and sampling rate to minimize local region’s underexposure, overexposure and motion blurring. Exposure control is implemented using all-integrated electronics without external optical modulation. This reduces overall system size and power consumption. 

The image sensor is implemented using a standard 130nm CMOS process while the exposure controller is implemented on a computer. We performed experiments under complex lighting and motion condition to test performance of the system, and demonstrate the benefit of pixel-wise adaptive imaging on the performance of computer vision tasks such as segmentation, motion estimation and object recognition.

\end{abstract}


%
\IEEEpeerreviewmaketitle

\section{Introduction}

CMOS image (CI) sensors are essential parts of computer vision (CV) systems in  autonomous robots, self-driving vehicles, and surveillance systems. Currently, CI sensors used for these systems are frame based with fixed global exposure that locks to the pixel readout timing. The inability to adjust each pixel’s exposure and frame rate independently can lead to regions of motion blur and overexposure.  Some examples are shown in Fig. 1(a)-(d) with regions of sub-optimal exposure settings. These saturated and/or blurred video frames cause information loss and lead to errors in a range of CV tasks such as motion estimation, object segmentation, and recognition. 

In this work, we demonstrate a camera system that can adaptively control each pixel’s exposure and sampling rate in real-time to maximize information content acquired from the scene. It does this using all electronic control without external  optical modulators. The system consists of three parts: an image sensor with pixel-wise exposure configurability, a high speed bi-directional chip to computer interface using the PCIe bus, and real-time a exposure controller implemented on a computer. 

The paper is organized as follows: In section II, we discuss previous work in the area of high dynamic range (HDR) and coded exposure imaging. In section III, we  describe the adaptive pixel-wise imaging system in detail. In section IV, we characterize adaptive exposure control using our system with experiment results. In section V, we address some areas of improvement in discussion and conclude the paper.

\begin{figure}
	\centering
	\includegraphics[width=\columnwidth]{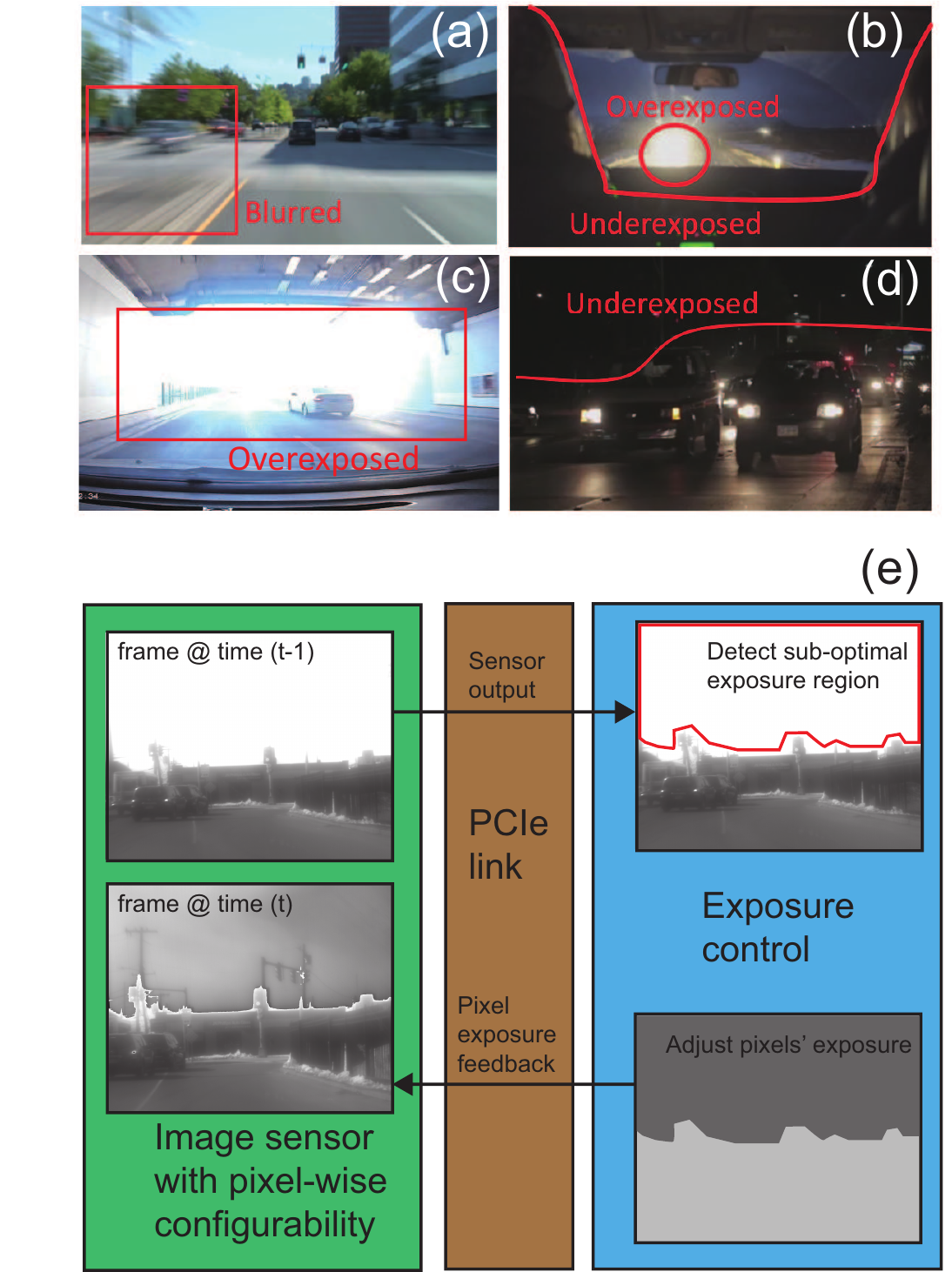}
	\caption{(a-d) Snapshot of videos acquired with frame based CI sensors. Highlighted area in the image are not optimally exposed. (a) Motion blur due to low frame rate. (b-d) Under and overexposure in image local regions. (a-d) are snapshots of videos from Shutterstock (www.shutterstock.com) (e) Block diagram of our system. It consists of three parts: an image sensor with pixel-wise exposure configurability, a high speed bi-directional chip to computer interface using the PCIe bus, and real-time exposure control algorithms.}
	\label{Fig1}
\end{figure}

\section{Summary of previous work}

Several previous works provide non-integrated technologies for adaptively controlling exposure patterns on conventional CI sensors. Here, we review each class of these works: 

\textit{Temporally varying exposure:} To minimize imaging imperfection due to suboptimal exposure, a conventional CI sensor can be used to sequentially capture multiple images of the same scene at different global exposures \cite{burt1993enhanced} \cite{mitsunaga1999radiometric}\cite{mann1994beingundigital}. Short exposure images capture details in bright regions while long exposure images enhance the signal quality in the dark regions. These sequential snapshots are then merged through tone-mapping algorithms to enhance overall scene dynamic range. While this method is widely used in photography, it is not suitable for non-stationary scenes as any movements during the sequential snapshots would cause ghosting in the tone-mapped HDR image. Furthermore, the need for multiple exposures limits the frame rate of the CI sensor, making it sub-optimal for video applications. 

\textit{Multiple image sensors:} We can extend the above methods to videos by using a multi-sensor system to capture the scene at different exposure settings simultaneously \cite{tocci2011versatile}\cite{aggarwal2004split}. These setups use beam splitters after the objective lens to generate multiple copies of the scene and then project them onto multiple detectors set at different exposures. While this method produces HDR frames in real-time, it requires multiple cameras and well-tuned optics for alignment. Further, spitting the image between two sensors attenuates collected light which degrades the maximum frame rate. 

\textit{Spatially varying transmittance:} Spatially modulating pixel transmittance can be used to enhance dynamic range without using multiple detectors\cite{nayar2000high}\cite{schechner2001generalized}. These methods use off-chip spatial light modulators (SLMs) prior to the CI sensor to modulate incoming light intensity at each pixel. By placing an optical spatially varying transmittance mask across the focal plane, it allows local pixel groups to capture a wider dynamic range. But to generate the HDR scene, spatial low pass filtering is needed which results in loss of spatial resolution. In addition, this technique suffers from the added complexity of the off-chip SLM, which must be well aligned to each pixel. 

\textit{Adaptive optical light intensity modulation:} The optical transmittance mask can be coupled with a feedback control loop to modulate light intensity at pixel-level \cite{nayar2000high}\cite{nayar2003adaptive}. These adaptive systems first detect over-exposed and dark region within the scene. They then adjust incident pixel-wise light intensity within the region to optimize the dynamic range. This method improves scene dynamic range without sacrificing spatial resolution. But like other optically modulated methods, these systems require precise optical alignment. In addition, the image sensor's framerate is fixed. Thus, these systems cannot increase pixel sampling speed during short exposure because the exposure clock must be tuned to the dimmest part of the scene. 

\section{Pixel-wise adaptive imaging system}

Here we describe an adaptive imaging system that optimizes exposure and frame rate at the pixel level. It achieves pixel-wise exposure control without external optical modulation. Figure 1(e) illustrates this system, which consists of three parts: (1) an image sensor with flexible pixel-wise exposure configurability, (2) a high speed chip computer bi-directional PCIe data link, and (3) exposure controller implemented on the computer. These three elements operate in a closed-loop to sample, detect and optimize each pixel’s exposure and frame rate to enhance image quality. The PCIe link and the exposure controller could be further integrated to a system-on-chip (SoC) solution to further decrease overall system size and power. 

Figure 1(e) also shows an example of this system’s operation. The image sensor outputs frame 1 of a video at time (t-1). In this frame, the sky region of the scene is over-exposed which leads to saturation of important objects including poles and traffic lights. From this frame, the exposure controller then segments the region of sub-optimal exposure and re-calculates the optimal exposure for this region. It feeds back updated exposure to the image sensor, which then uses it to sample the next frame at time (t), where both the sky and the road regions of the scene are optimally sampled. In this section, we describe each part of the proposed system in detail. 

\subsection{Image sensor with pixel-wise exposure configurability}

A number of image sensors with pixel-wise exposure configurability have been demonstrated, including our previous work \cite{zhang2016compact}\cite{luo2017exposure}\cite{sarhangnejad20195}. We first proposed a pixel-wise configurable exposure image sensor architecture \cite{zhang2016compact}. However, as we shall discuss here in detail, this first design’s timing causes small unintended sampling deadtime during exposure. Subsequently, Luo, Ho and Mirabbasi, Sarhangnejad et al. have proposed dual-tap approaches which use two floating diffusions ($FD$) to collect charges from the photodiode\cite{luo2017exposure}\cite{sarhangnejad20195}. This design ensures minimal sampling deadtime but doubles the number of readout circuits. In this work, we improve on our previous design to ensure all the charges are accumulated during pixel exposure. This pixel can achieve the performance of dual-tap design without using additional readout circuits.
\newline

\subsubsection{Image sensor architecture} 

The image sensor design is outlined in Fig. 2. The pixel array consists of 256 $\times$ 256 pixels with 6.5$\mu m$ pixel pitch and is fabricated using a standard 130$nm$ CMOS process. Pixel circuitry is modified based on our previous work and consists of three parts: a 4T pixel, an exposure control ($EX$) gate and an in-pixel static random access memory (SRAM) block. The 4T pixel include a pinned-photodiode ($PD$), a transfer gate ($TX$), a reset gate ($RST$) and a row selection gate ($RSEL$) \cite{zhang2016compact}\cite{fossum2014review}. We inserted the EX gate between PD and the readout circuitry for pixel-wise exposure control. An in-pixel 1-bit static random access memory (SRAM) block controls the $EX$ gate to modulate pixel-wise exposure according to the timing diagram in Fig. 3 and Fig. 4. 

Unlike our previous design where $EX$ is placed between $PD$ and $TX$ \cite{zhang2016compact}, in this design the $EX$ gate is inserted between $TX$ and $FD$. As we shall elaborate in the next section, this modification fixes sampling deadtime caused by charge dissipation that was present in our previous design. 

On the chip level, we use analog correlated double sampling (CDS) and cyclic ADCs for column readout. Compared to successive approximation register (SAR) ADC used in our previous work, the cyclic ADC occupies less area by minimizing  the use of on-chip capacitors \cite{mase2005wide}\cite{park2009high}. The schematic and timing of the ADC are shown in Fig. 3. The operational transconductance amplifier (OTA) is shared between the CDS circuit and ADC.  We choose a single-ended architecture over the fully differential one to further save chip area. Each cyclic ADC’s layout has dimension of 6.5$\mu m \times$  200$\mu m$ to fit into the pixel column pitch. 

Pixel-wise exposure control is done through the SRAM drivers located at the bottom of each column. When a row is selected for readout through $RSEL$, these SRAM drivers are  loaded with the exposure bits for all the pixel within that row. The timing for SRAM writing procedure is explained in the next subsection. 

\begin{figure}
	\centering
	\includegraphics[width=\columnwidth]{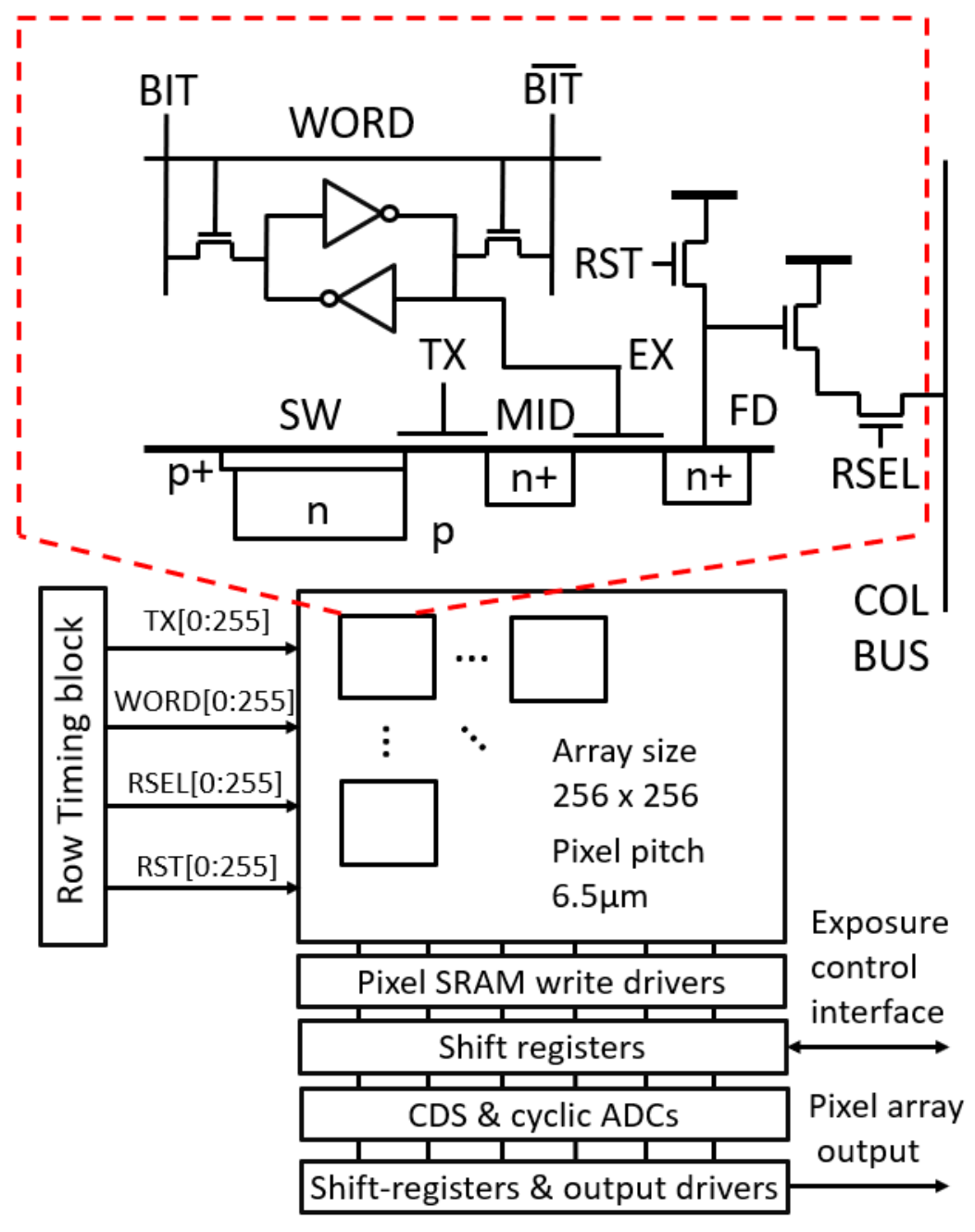}
	\caption{Pixel-wise image sensor architecture. The sensor is fabricated in a standard 130nm CMOS process featuring an array of 256 $\times$ 256 pixels with 6.5$\mu m$ pitch}
	\label{Fig2}
\end{figure} 

\subsubsection{Timing} 

The circuit timing diagram is shown in Fig. 3 for a pixel located at row zero and column one. When this row is selected for readout, we pre-charge the $BIT$ line for this column ($BIT[1]$) to high, indicating this pixel is ending its exposure. The CDS circuit then samples the reset value onto $C_1$, while putting voltage $V_{RP}-V_{COM}$ across $C_2$. The $WORD$ line for row zero ($WORD[0]$) subsequently pulses to write the SRAM with the logic value of $BIT$. This turns $EX[0,1]$ gate ON, connecting $PD$ to the readout structure. Then $TX$ pulses to transfer light generated charge from the $PD$ onto the $FD$ node. $FD$ voltage is subsequently sampled onto $C_1$ while $C_2$ is place in the feedback path of the amplifier. At the end of the CDS signal phase, the voltage, $V_{OUT}$, is:

\begin{equation} 
V_{OUT} = \frac{C_1}{C_2}(V_{RESET} - V_{SIGNAL}) + V_{RP}
\end{equation}

where $V_{RESET}$ is the $FD$ reset value and $V_{SIGNAL}$ is $FD$ signal value after charge transfer.  $V_{RP}$ is the voltage to adjust $V_{OUT}$ into the proper ADC dynamic range. Note that the CDS’s reset sampling must occur before EX is programmed to release charge onto the $FD$ node. 

Analog to digital (AD) conversion begins after the conclusion of CDS. We divide the digitization process into two repeating sub-phases driven by two non-overlapping clocks ($\phi_1$ and $\phi_2$). During the sampling sub-phase, the ADC samples the output of the OTA onto the capacitor $C_1$ and $C_2$. Then during the amplification phase, it drives $V_{OUT}$ to the value given by:

\begin{equation} 
V_{OUT} = 2V_{OTA} - V_R
\end{equation}

where $V_R$ is determined by the 1.5-bit MDAC based on the output of the sub-ADC:   

\begin{equation} 
V_R = 
\begin{cases}
    V_{RH},& \text{if } D = 2\\
    V_{COM},& \text{if } D = 1\\
    V_{RL}, & \text{if } D = 0
\end{cases}
\end{equation}

The MDAC generates 1.5-bit per conversion stage. This is a widely used technique in pipeline/cyclic ADC design to relax sub-ADC’s comparators precision \cite{mase2005wide}\cite{park2009high}. The ADC repeats sampling and amplification sub-phase ten times. Addition logic then generates 10-bit digital output based on each stage’s results and transmit the digital word to a serial interface for readout. 

\begin{figure}
	\centering
	\includegraphics[width=\columnwidth]{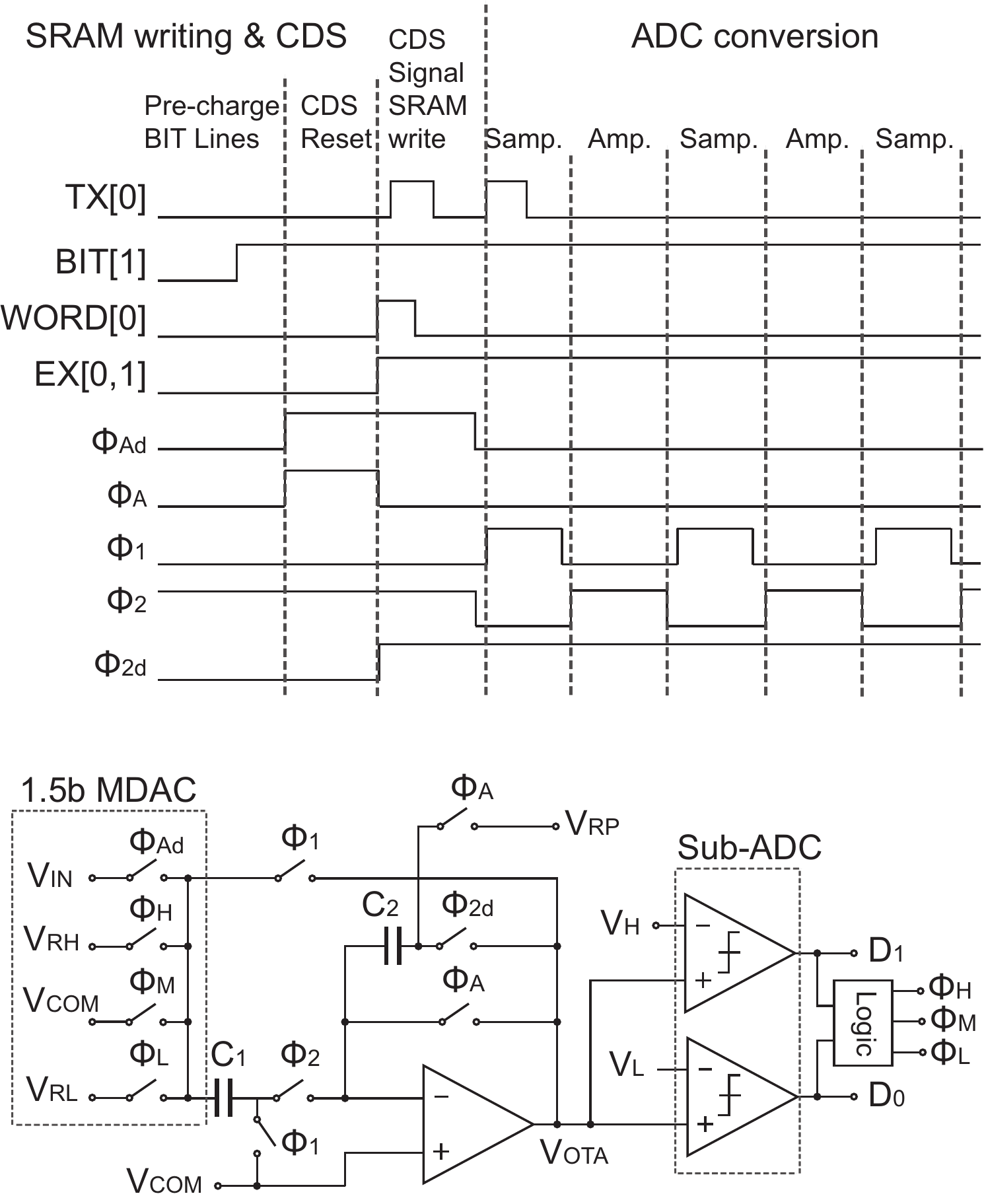}
	\caption{An example pixel timing diagram and the schematic of the CDS and column cyclic ADC circuits}
	\label{Fig3}
\end{figure} 

\subsubsection{Modified pixel design}

Fig. 4(a) shows the pixel timing across multiple frames and charge potential diagrams to illustrate charge transferring at key timing point. We use this figure to illustrate the sampling deadtime present in our previous design, and our design modification.  

To end this pixel’s previous exposure, we pull $EX$ signal high to sample the pixel. This pixel potential is then reset at timing point A. The $EX$ remains high until this pixel is selected at the next frame at timing point B. In our previous design \cite{zhang2016compact}, where $EX$ is inserted between $PD$ and $TX$, the photo-generated charge will pour into $MID$ when $EX$ is high between timing A and B (Fig. 4(b)). These charges would then be trapped at $MID$ when $EX$ is pulled low at point B, and subsequently cleared when $TX$ and $RST$ are pulled high at timing point C. Therefore, these charges would be absent when the pixel value is readout at the end of exposure (timing point D). With this charge dissipation, the final charge measured at point D is the charges collected between B and D and we lose the charges collected between A and B. Of course, since $MID$’s well-potential is much smaller compare to FD and PD, this non-ideality can be negligible in bright light conditions. But it would nevertheless be a problem when overall signal is small.

Aiming to fix this issue, we switch the location of $EX$ and $TX$ gate such that $TX$ is placed between $PD$ and $EX$ (Fig. 4(c)). Between timing point A and B, the charges are kept at $PD$ since $TX$ is always high. At timing point B, $EX$ is pulled high before $TX$ to rise the potential barrier. The subsequent TX pulses would transfer charge onto the $MID$ node. But $MID$ node would be protected from reset by the $EX$ potential barrier. At timing location D, $EX$ is lowered to allow this pixel to be sampled. Note that during CDS, we sample the reset prior to lowering $EX$. Therefore, the charge trapped at $MID$ would be counted as the signal voltage. The resulting sample at point D would the charges collected between point A and D.  

\begin{figure}
	\centering
	\includegraphics[width=\columnwidth]{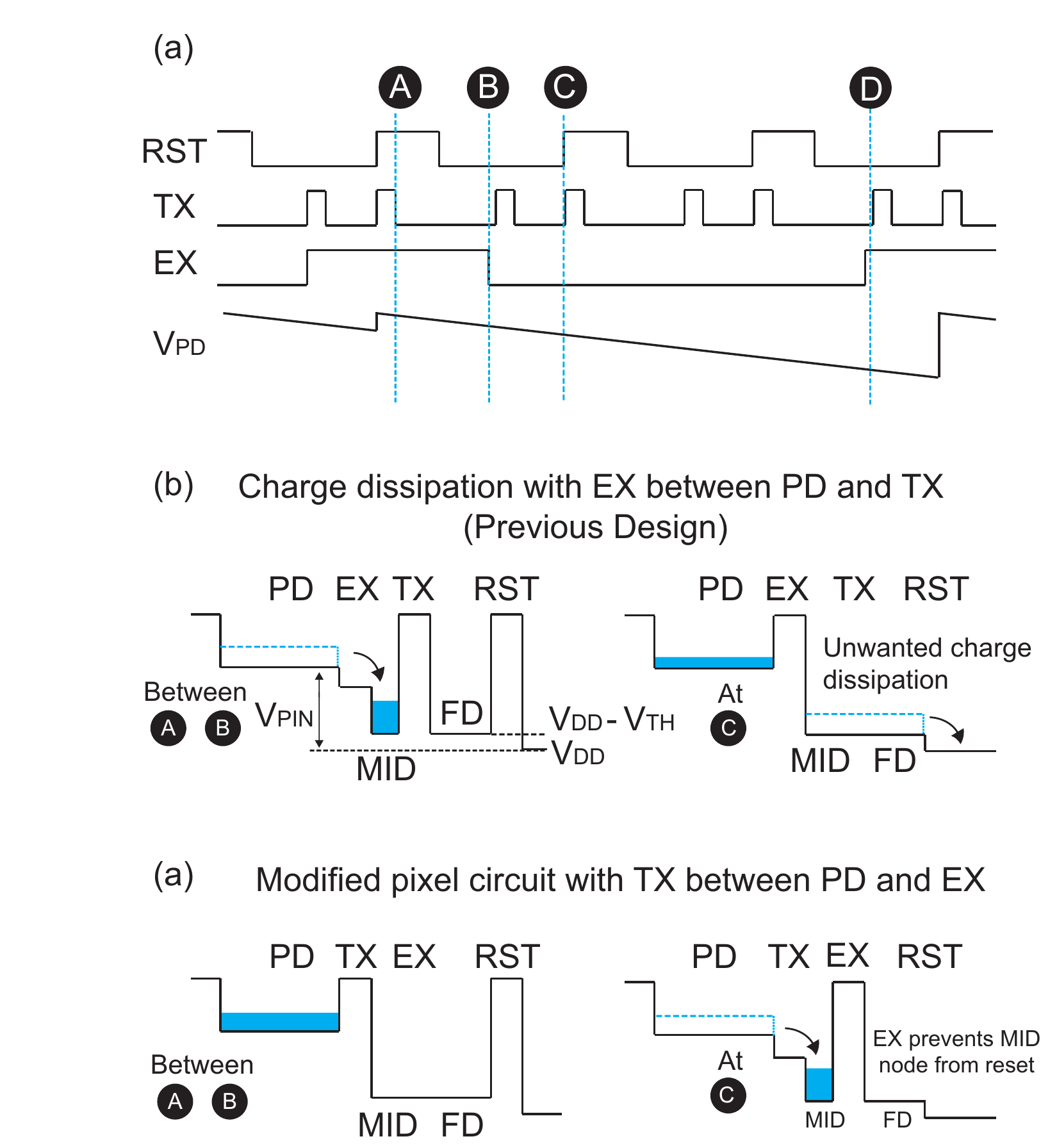}
	\caption{Timing diagram during pixel pixel exposure and pixel charge transfer (a) pixel timing diagram across multiple frames (b) charge transfer plot of pixels in our previous design \cite{zhang2016compact} (b) charge transfer plot of modified pixels in this work }
	\label{Fig5}
\end{figure}

\subsection{Chip-computer bi-directional interface }

To prototype pixel exposure control algorithms, it is convenient to work with a high level language running on a desktop PC. However, with a computer in the loop it is imperative to minimize communication delay times to prevent closed-loop instability. To achieve single frame delay in the exposure control, we must minimize the communication delay from the chip to computer so more time can be allocated to the exposure control algorithms on the computer side. We have selected the PCIe bus for chip-computer communication due to its high bandwidth and low transmission latency. We implement the bi-directional PCIe interface using a Xilinx Kintex-7 FPGA with IP cores from Xillybus \cite{xillybus.com}. We characterized the round-time communication delay to average around 70$\mu s$, with maximum delay of 85$\mu s$ \cite{oe2016}. This makes communication speed negligible compared to the time between frame (for framerate $<$100 FPS). 

In addition to hosting the bi-directional PCIe interface, the same Kintex-7 FPGA also implements logic to convert pixels' exposure values from the computer to serial data streams. These data streams are then send to the image sensor to load the shift registers at the input of the SRAM drivers, shown in Fig. 2.

To rapidly prototype different feedback control algorithms, we made use of the Bonsai visual programming language \cite{lopes2015bonsai}. We developed Bonsai plugins on top of a data streaming API \cite{jonathan_newman_2019_3254431}. Specifically, we developed a source plugin (PCECameraRead) to capture images from the camera and a sink (PCECameraWrite) to write pixel exposure times to the camera using the bi-directional PCIe link. Code for these plugins is available on the Bonsai.CodedPixel git repository (https://github.com/jonnew/Bonsai.CodedPixel). Bonsai includes many high performance image processing components and easy to use stream synchronization elements making it ideal for our purposes. Bonsai workflows implementing for each of the control algorithms presented in this paper can be found in the algorithms folder of the CodedPixel git repository.

\subsection{Pixel-wise exposure and sample rate controller}

For prototyping, the pixel-wise exposure controller is hosted on a computer linked to the image sensor using the PCIe interface. By controlling pixel-wise exposure, the controller can also adjust each pixel’s sampling rate. The goal of the controller is to optimize pixel exposure based on estimates of local intensity and motion information. Pixel exposure should be adjusted to avoid overexposure, underexposure and motion blur. We divide this optimization task into two control modes: exposure optimization based on (1) intensity and (2) local scene motion.	

\subsubsection{PI mode: Exposure optimization based on intensity}

We used a proportional integral (PI) controller to control pixel exposure based on its intensity measurement. The goal is to maintain the intensity of each pixel,  $I_{x,y}$ within a desired range that is set just below the point of over-exposure, similar to \cite{nayar2003adaptive}. To define this controller, we first calculate an intensity error for every pixel:

\begin{equation} 
\epsilon_{x,y} = 
\begin{cases}
    I_{x,y}(n)-I_{target},& \text{if } |I_{x,y} -I_{target}| > e_{tol} \\
    0, & \text{otherwise }
\end{cases}
\end{equation}

where $\epsilon_{x,y}(n)$ is the intensity error for pixel at row $x$ and column $y$. $I_{target}$ is a constant representing the desired intensity. The positive constant,  $e_{tol}$, is an error tolerance such that $\epsilon_{x,y}(n)$ is non-zero only when  $|I_{x,y}-I_{target}|$ is larger than $e_{tol}$. With the error term defined, the PI controller is written as:

\begin{equation} 
r_{x,y}(n) = K_p\epsilon_{x,y}(n) + K_i\sum_{n'=0}^{n}\epsilon_{x,y}(n')
\end{equation}

\begin{equation} 
E_{x,y}(n) = \lfloor E_{x,y}(n-1) - r_{x,y}(n) \rfloor
\end{equation}

where $r_{x,y}(n)$ is the output of the PI controller. $K_p$ and $K_i$ are the proportional and integral gain respectively. Finally, the pixel exposure value, $E_{x,y}(n)$, is an integer representing the multiples of the shortest exposure. We calculate the $E_{x,y}(n)$ as the numerical floor of the difference between the previous exposure $E_{x,y}(n)$ and $r_{x,y}(n)$. the PI controller will minimize $r_{x,y}(n)$ to achieve steady state $E_{x,y}(n)$ and $I_{x,y} \approx I_{target}$.

There are four parameters to set for the PI controller: $I_{target}$, $e_{tol}$, $K_p$ and $K_i$. The choice of these parameters depends on the pixel intensity range and the max allowable $E_{x,y}(n)$. For example, if $1 \leq E_{x,y}(n) \leq 8$ and in a standard image format where pixel value range from 0 – 255, $I_{target}$ can be selected to be around 200, with $e_{tol}$ of 30. That means the PI controller will try to keep optimal pixel in the range between 170 to 230. We should increase $K_p$  to speed up settling speed. But its value should not be too larger to induce instability. $K_i$ should be kept as a small fraction of $K_p$ to eliminate residual error after the application of the proportional control. We select these gain parameters using the following rule of thumb: 

\begin{equation} 
K_p \textless \frac{max(E_{x,y})}{max(\epsilon_{x,y})} \text{  and  } K_i \textless 0.1 \dot K_p
\end{equation}

where $K_p$ should be less than the quotient of maximum exposure and maximum intensity error, and  $K_i$ should be less than 10\% of $K_p$.  This parameter selection ensures that in most extreme condition $r_{x,y}$ does not cause $E_{x,y}$ to overshoot and create instability. In this example, if $I_{target}$ is 200, then $K_p$ should be less than $8/170 = 0.04$ and $K_i$ should be less than 0.004.

\subsubsection{OF mode: Exposure optimization based on motion}

For a static scene, it is optimal to use longer pixel exposure and slower frame rate to enhance SNR. But for motion scenes, it is desirable to use short pixel exposure and fast frame rate to minimize motion blur and to avoid temporal aliasing. Because the scene can change during camera operation,  using a  derivative (D) term to estimate temporal trends in pixel exposure is ineffective (e.g. in the case of a sharp edge or line passing through the scene). Instead, we update pixel exposure by directly estimating local motion using optical flow (OF) measurements \cite{farneback2003two}\cite{lucas1981iterative}.  This measurement can be used to predicts scene motion which is used to preemptively tune pixel exposure.

However, using OF-based control introduces some issues into the PI controller. First, for a static scene, when we change exposure at $E(n-1)$ using the PI controller, it results in non-zero optical flow vectors computed at time $t$ due to the pixel intensity change. This “false” OF value is representative of our control signal and not actual scene motion and therefore causes an error on motion driven exposure optimization. Second, OF measurement can be noisy at the pixel level, especially at low light condition with low SNR. Further, OF measurements' accuracy can also be affected by motion blur caused by long exposures. All of these factors make pixel-wise motion driven exposure difficult to achieve. 

To address these issues, we allow the PI controller and the OF controller to engage pixels separately depending on the magnitude of pixel motion. We first divide the image into sub-blocks of size $M \times M$ pixels. We then calculate the average OF for all the pixels within each of these blocks over a period of $T_{switch}$:

\begin{equation} 
v_{i,j}(n) = \frac{1}{M^2\dot T_{switch}} \sum_{y=1}^{M} \sum_{x=1}^{M} \sum_{n'=n-T_{switch}}^{n}|u_{x,y}(n')|
\end{equation}

where $|u_{x,y} |$ is magnitude of the optical flow vector at pixel $(x,y)$. $v_{i,j}$ is the average of $|u_{x,y}|$ over a period $T_{switch}$ for all the pixels within this block. Spatially averaging of OF enhances measurement accuracy, while temporal averaging over a period of $T_{switch}$ allows us to distinguish motion from instantaneous change in pixel value due to closed-loop exposure update. After incorporating  $v_{i,j}(n)$, pixel exposures are determined as:

\begin{equation} 
E_{x,y}(n) = 
\begin{cases}
    E_{PI_{x,y}}(n),& \text{if } v_{i,j} < v_{tol}  \text{\; \; PI mode} \\ 
    \lfloor max(E_{x,y})-K_v v_{i,j} \rfloor, & \text{if } v_{i,j} \geq v_{tol} \text{\; \; OF mode}
\end{cases}
\end{equation}
  
$v_{tol}$ is a constant threshold to determine controller mode selection. For static regions or regions with instantaneous motion, we allow PI mode to control pixel exposure. But for regions with constant motion, we use OF mode which calculates exposure based on the magnitude of the motion. $K_v$ is the gain that converts the OF magnitude, measured in units of pixels moved, to exposure values. $K_v$ was determined using test images, where lines are moved in front of the camera at different speed to determine the correct $K_v$ to avoid motion blurring. 

\section{Experiments}

We next experimentally investigated the adaptive exposure feedback imaging system. In all the experiments described here, $E_{x,y}$ is represented by a 3-bit integer. $E_{x,y}=1$ corresponds to shortest pixel exposure of $30$ ms, while longest exposure $E_{x,y}=8$ corresponds to pixel exposure of $240$ ms. This setting also configures the fastest pixel sampling rate to around 30 frames per second (FPS) and slowest sampling rate to around 4 FPS. Before the raw images from the sensor is sent to the exposure controllers, they first undergo fixed-pattern noise correction and are filtered by a 3 $\times$ 3 median filter to remove salt-and-pepper noise from non-responsive pixels.

In these experiments, we take the 10-bit ADC full scale as pixel's value range $(0 – 1024)$. We set the PI controller parameter following equation (7): $I_{target} = 800$, $\epsilon_{tor} = 120$, $K_p = 0.01$ and $K_i = 0.001$. We compute the OF measurements using the Farnelback method based on polynomial expansion\cite{farneback2003two}. We set $K_v = 2$, this means for every motion magnitude $v_{x,y}$ of 1 pixels, we reduce exposure, $E_{x,y}$, by 2 steps. $M$ is set to be 8 corresponding to $v_{x,y}$ averaging over an 8 $\times$ 8 blocks. 

\subsection{Setup}

Fig. 6 shows the chip micrograph, the data collection setup, and two scenes used for experiments. The chip is wirebonded onto the PCB and mounted behind an objective lens, shown in Fig. 6 (a) and (d). The data from the image sensor is transmitted to a connector board which sends the data into the KC705 FPGA its FPGA Mezzanine Card (FMC)  port, shown in Fig. 6(b). The KC705 hosts the PCIe bi-directional interface and is mounted onto one of computer’s PCIe slots. Fig. 6 (c) is a scene used in this experiment. It is a cardboard cutout in a shape of a cartoon sun. We overlay it with a transparent paper with a drawing of a smiley face. This setup is then back-illuminated to create high dynamic range scene for imaging experiment. Fig. 6 (d) is a setup to create continuous motion to test the functionality of the controller in OF mode. Objects are mounted on a rotating stand. The spinning motion then creates continuous motion at the top-half of the video, while the bottom half remain stationary. The detailed experiment results are described in the next sections. 

\begin{figure}
	\centering
	\includegraphics[width=\columnwidth]{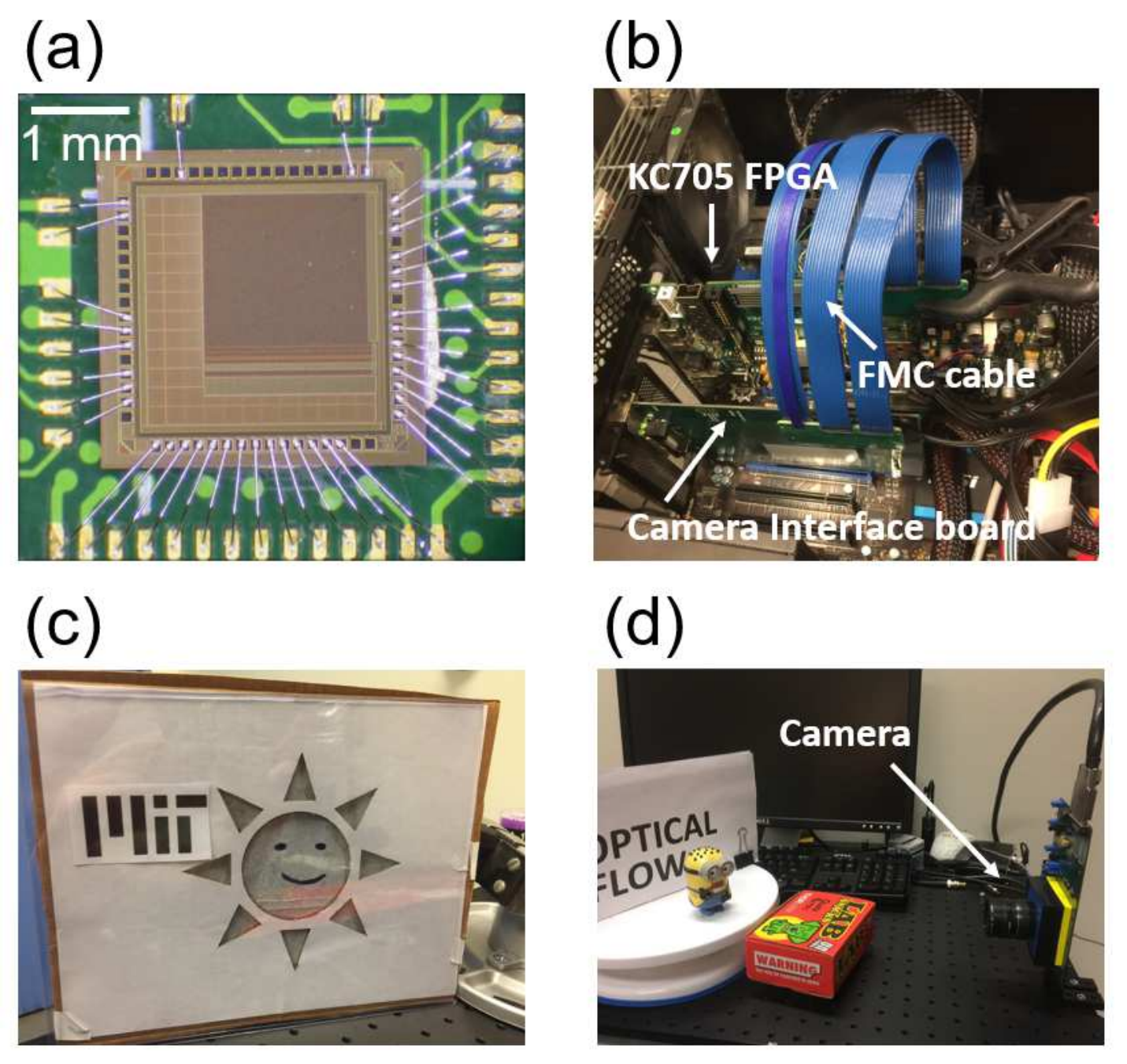}
	\caption{Chip micrograph and experimental setups: (a) chip micrograph (b) data collection setup (c and d) scenes and objects used for testing the PI controller (c) and OF controller (d)}
	\label{Fig6}
\end{figure} 

\subsection{Stationary scenes}

Fig. 6 shows a back-illuminated pictures of a cartoon cardboard cutout in the shape of a Sun under different exposure settings. In Fig. 6(a) and (b), the entire scene undergoes short (30ms) and long (240ms) global exposures respectively. During short exposure, the smiley face overlaid on top of the circle, and the gap between the circle and the triangles cutouts are visible and can be easily segmented. But the background remains dark. Conversely, during long exposure, the background text of the MIT logo is visible, but the circle and the triangles of the Sun are over-exposed. As a result, the edges between the triangle and the circle shrinks due to over-exposure. This makes the segmentation more difficult. In addition, due to over-exposure, the smiley face is statured and not visible.

Fig. 6(c) shows an example when each pixel’s exposure is tuned based on its intensity using the pixel-wise exposure controller as described in section III.C. Since the camera is imaging a static scene, the control adjusts pixel-wise exposure in PI mode. The exposure pattern to needed to generate Fig. 6(c) is shown in Fig. 6(d). The controller sets most of the background to long exposure of 240ms to enhance the SNR. It only lowers the exposure values for pixels within the back-illuminated circle and triangles to avoid over-exposure. The pixels around the bright regions are tuned to have exposure between 30ms and 240ms. As a result, both the background text and the smiley face at the back-illuminated region are visible. The gaps between the circle and triangles are also wide and distinguishable.

\begin{figure}
	\centering
	\includegraphics[width=\columnwidth]{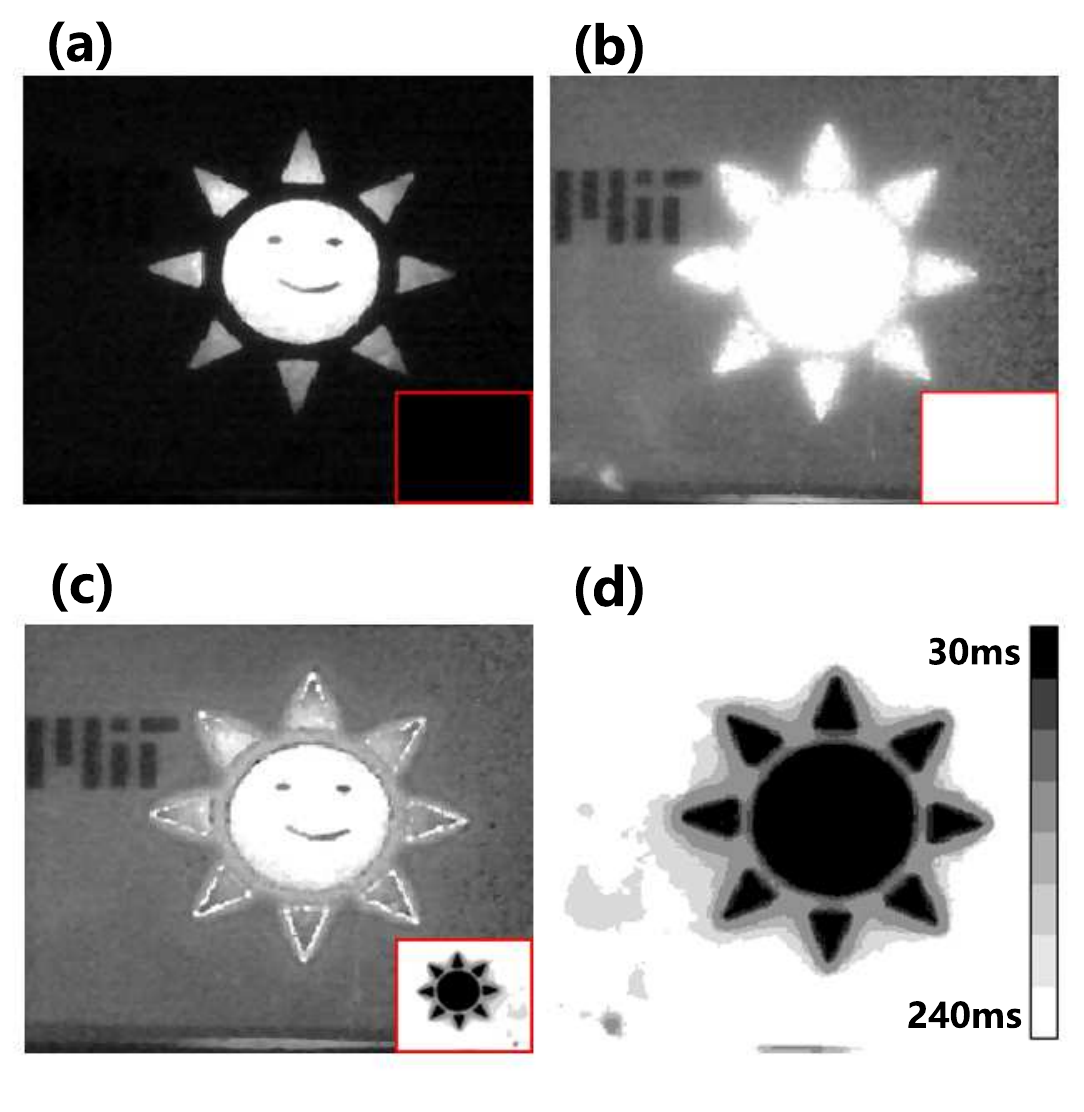}
	\caption{Pixel-wise exposure optimization for stationary scene using controller in PI mode. (a) Image acquire with short global exposure, (b)  long global exposure. (c) Image acquired with optimized pixel-wise exposure set by the PI controller}
	\label{Fig6}
\end{figure} 

\begin{figure*}
	\centering
	\includegraphics[width=\textwidth]{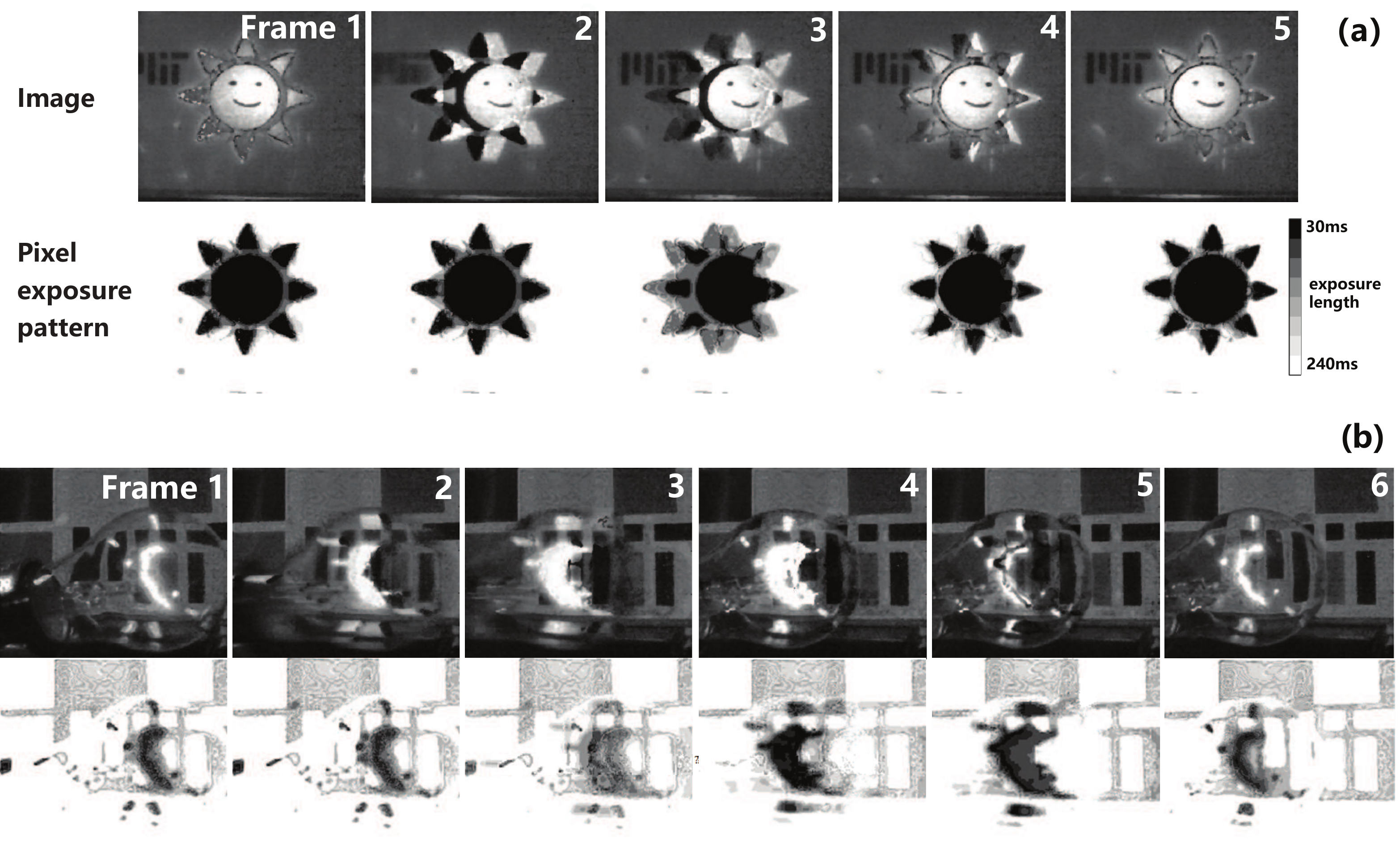}
	\caption{Pixel-wise exposure optimization for scenes with instantaneous motion using controller in PI mode. Two video examples are shown in (a) and (b). The top row of each example shows the image sensor output, while the bottom row shows the corresponding pixel exposure pattern. The legend for pixel exposure length is also}
	\label{Fig7}
\end{figure*} 

\begin{figure*}
    \centering
    \includegraphics[width=\textwidth]{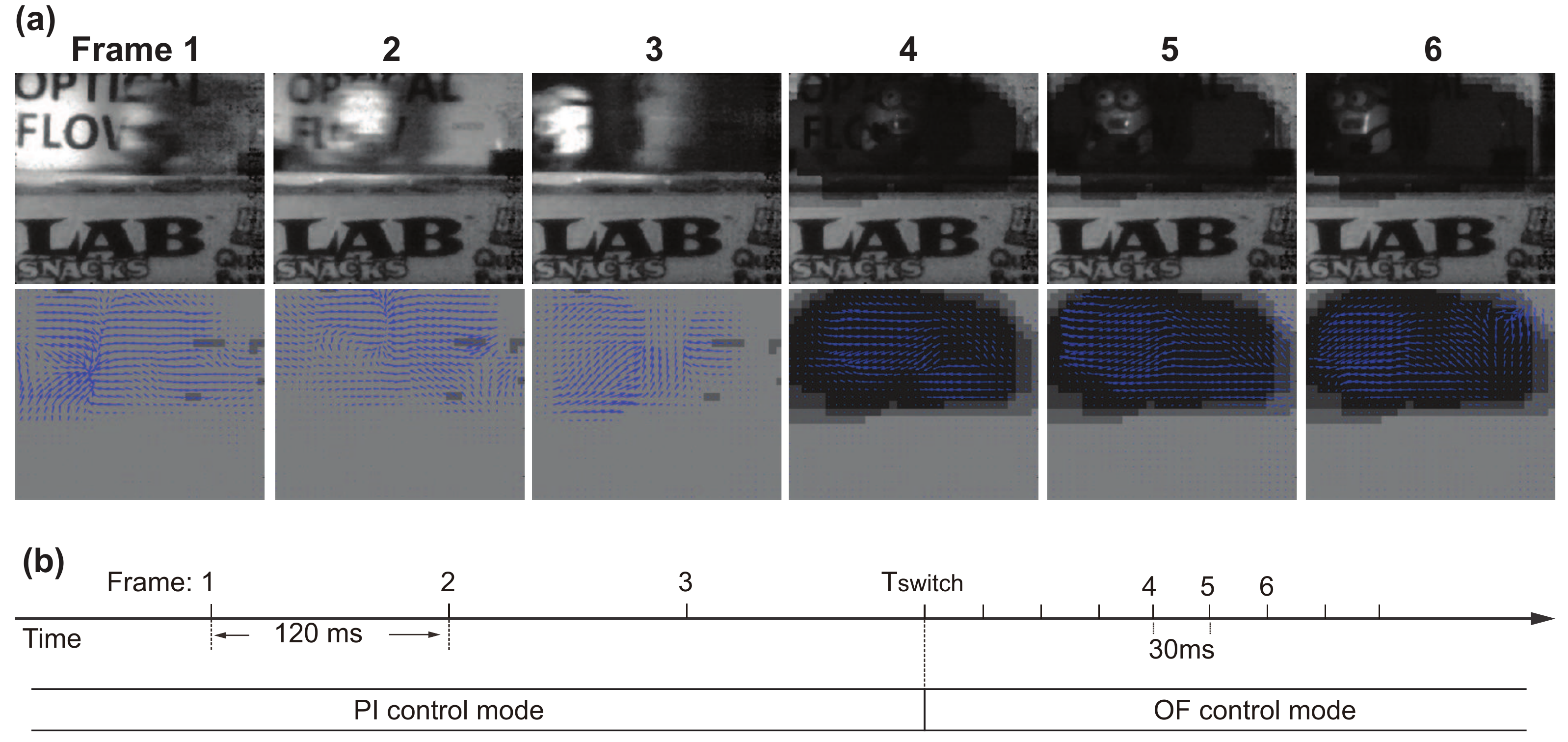}
    \caption{Pixel-wise exposure optimization for scenes with continuous motion using controller in OF mode. (a) Six frame from a video where image sensor outputs are shown in the top row, while pixel exposure patterns and OF measurements are shown in the bottom row. (b) A timeline showing the duration between each frames. when pixels exposure are shorten at $T_{switch}$, the pixel sampling rate also increase in the corresponding region}
    \label{classification}
\end{figure*}

\subsection{Motion scenes}

\textit{Instantaneous motion:} Fig. 7 demonstrates the controller’s performance for scenes with instantaneous motion. We show two examples here. For each example, we plot the image sensor output on the top row and the corresponding pixel exposure pattern is on the bottom row. In the example (a), we move the back-illuminated cardboard cutout to the right at frame 2. This motion is not large enough to cause the controller to switch to OF mode from PI mode. As a result, pixels in the region where the motion occurs become overexposed and underexposed. This causes the error function to grow and PI controller starts to regulate pixels exposure, shown in frame 3 and 4. The pixels’ exposure eventually settles at frame 5, after 3 frames following the end of motion 

Fig. 7(b) shows another example, where an incandescent light bulb is placed in front of an MIT logo. The PI controller tunes the pixels at bulb’s filament to short exposure while keeping the background at longer exposure. This enables us to see both the shape of the filament and the background. We then introduce motion at frame 2 and 3 by moving the bulb to the right. Pixels at the motion region become sub-optimally exposed. But PI controller then optimizes the them which eventually settles to their new exposure at frame 6.

\textit{Continuous motion:} Fig. 8 demonstrates an example where continuous motion triggers the OF mode for pixel exposure control. In this example, objects are placed on a rotating stand, as shown in Fig. 6(c). We then spin the stand to generate continuous motion. This results in continuous object motion on the top section of the video frame, while the bottom section remain stationary.

Fig. 8(a) top row shows the outputs of image sensors, while the bottom row shows pixel exposure and corresponding OF vectors. Initially, each pixel is configured to have exposure of 120ms, sampling at around 8 FPS. Some example image are shown in frame 1 – 3. The long pixel exposures, although optimal from SNR perspective, generate blurring effects in frame 1 to 3. 

The optical flow measurements are continuously integrated according to equation (8) and It reaches the threshold $v_{tol}$ at end of period, $T_{switch}$. This causes the controller to switch from PI mode to OF mode according to equation (9). In OF mode, the pixel exposures are optimized according to the magnitude of the velocity instead of intensity. In this example, it shortens pixel exposure and increase framerate for the top region of the frame for the amount proportional to the magnitude of the motion. These pixels now update every 30ms instead of 120ms and results in blur-less capturing of the motion. 

It can also be observed that the OF measurements, especially the direction of the motion vectors in frame 1-3 are less accurate in reflecting the actual motion due to the blurring effects from long exposures. This is compared to OF measurements in frame 4-6 during where blurless frames results in much more accurate motion measurements.

\section{Discussion, future work and conclusion}

We have demonstrated a closed-loop, all-electronic, pixel-wise exposure control system that is capable of adapting to both changes in scene intensity and motion in order to tune exposure in over-and-over exposed scene regions. We have provided a simple PI control algorithm to tune each pixels exposure time in order to achieve ultra-high dynamic range imaging of static scenes with large differential lighting levels. Further, we have provided exposure control based on predicted scene motion using optical flow in order to adjust pixel exposure times to minimize blurring.

Our current system uses a PCIe based communication protocol and high-level software in the control loop for prototyping purposes. In the future, we foresee the complete integration of the control logic system directly into the camera sensor to form an SoC capable of native and automatic adaptive pixel exposure control. In addition, several areas of the system could also be improved in future works: 

\subsubsection{Application dependent metrics for measuring sub-optimally sampled pixels}

The key to high performance controller is well-defined error functions. In our controller, we used intensity error to measure over and under exposure, and in addition used optical flow as an indicator for motion blur. These performance metrics are selected for general applications. But a better approach would be to find application dependent metric to measure sub-optimally sampled pixels. For example, when the videos are inputs to classification or segmentation tasks, we could use classification or segmentation accuracy as a metric to adjust exposure and frame rate.

\subsubsection{Combining learning to pixel-wise exposure estimation}

The recent development in convolutional neural network (CNN) could also be utilized to learn the optimal pixel exposure patterns. Previous work using CNN for optical flow estimation and motion blur detection and removal have been proposed \cite{dosovitskiy2015flownet}\cite{sun2015learning}\cite{nah2017deep}. A similar network architecture could be used to segment the scene based optical pixel exposure. A potential difficulty with this approach is the collection of high dynamic range and multi-frame rate video data for network training.   

\subsubsection{Using coded exposure patterns}

The most difficult case for the exposure controller is when there is high continuous motion with low illumination. Prolonging exposure increases SNR but leads to motion blur. Shortening exposure reduces motion blur but causes low SNR. In this case, we could configure the pixels to sample in coded exposure patterns and use computational imaging methods to enhance image quality [9], [22].

\bibliographystyle{IEEEtran}
\bibliography{references}

\begin{thebibliography}{10}
\providecommand{\url}[1]{#1}
\csname url@samestyle\endcsname
\providecommand{\newblock}{\relax}
\providecommand{\bibinfo}[2]{#2}
\providecommand{\BIBentrySTDinterwordspacing}{\spaceskip=0pt\relax}
\providecommand{\BIBentryALTinterwordstretchfactor}{4}
\providecommand{\BIBentryALTinterwordspacing}{\spaceskip=\fontdimen2\font plus
\BIBentryALTinterwordstretchfactor\fontdimen3\font minus
  \fontdimen4\font\relax}
\providecommand{\BIBforeignlanguage}[2]{{%
\expandafter\ifx\csname l@#1\endcsname\relax
\typeout{** WARNING: IEEEtran.bst: No hyphenation pattern has been}%
\typeout{** loaded for the language `#1'. Using the pattern for}%
\typeout{** the default language instead.}%
\else
\language=\csname l@#1\endcsname
\fi
#2}}
\providecommand{\BIBdecl}{\relax}
\BIBdecl

\bibitem{burt1993enhanced}
P.~J. Burt and R.~J. Kolczynski, ``Enhanced image capture through fusion,'' in
  \emph{1993 (4th) International Conference on Computer Vision}.\hskip 1em plus
  0.5em minus 0.4em\relax IEEE, 1993, pp. 173--182.

\bibitem{mitsunaga1999radiometric}
T.~Mitsunaga and S.~K. Nayar, ``Radiometric self calibration,'' in
  \emph{Proceedings. 1999 IEEE Computer Society Conference on Computer Vision
  and Pattern Recognition (Cat. No PR00149)}, vol.~1.\hskip 1em plus 0.5em
  minus 0.4em\relax IEEE, 1999, pp. 374--380.

\bibitem{mann1994beingundigital}
S.~Mann and R.~Picard, ``On being `undigital' with digital camera: extending
  dynamic range by combining differently exposed pictures,'' \emph{MIT Media
  Lab Perceptual}, vol.~1, p.~2, 1994.

\bibitem{tocci2011versatile}
M.~D. Tocci, C.~Kiser, N.~Tocci, and P.~Sen, ``A versatile hdr video production
  system,'' in \emph{ACM Transactions on Graphics (TOG)}, vol.~30, no.~4.\hskip
  1em plus 0.5em minus 0.4em\relax ACM, 2011, p.~41.

\bibitem{aggarwal2004split}
M.~Aggarwal and N.~Ahuja, ``Split aperture imaging for high dynamic range,''
  \emph{International Journal of Computer Vision}, vol.~58, no.~1, pp. 7--17,
  2004.

\bibitem{nayar2000high}
S.~K. Nayar and T.~Mitsunaga, ``High dynamic range imaging: Spatially varying
  pixel exposures,'' in \emph{Computer Vision and Pattern Recognition, 2000.
  Proceedings. IEEE Conference on}, vol.~1.\hskip 1em plus 0.5em minus
  0.4em\relax IEEE, 2000, pp. 472--479.

\bibitem{schechner2001generalized}
Y.~Y. Schechner and S.~K. Nayar, ``Generalized mosaicing,'' in
  \emph{Proceedings Eighth IEEE International Conference on Computer Vision.
  ICCV 2001}, vol.~1.\hskip 1em plus 0.5em minus 0.4em\relax IEEE, 2001, pp.
  17--24.

\bibitem{nayar2003adaptive}
S.~K. Nayar and V.~Branzoi, ``Adaptive dynamic range imaging: Optical control
  of pixel exposures over space and time,'' in \emph{null}.\hskip 1em plus
  0.5em minus 0.4em\relax IEEE, 2003, p. 1168.

\bibitem{zhang2016compact}
J.~Zhang, T.~Xiong, T.~Tran, S.~Chin, and R.~Etienne-Cummings, ``Compact
  all-cmos spatiotemporal compressive sensing video camera with pixel-wise
  coded exposure,'' \emph{Optics express}, vol.~24, no.~8, pp. 9013--9024,
  2016.

\bibitem{luo2017exposure}
Y.~Luo, D.~Ho, and S.~Mirabbasi, ``Exposure-programmable cmos pixel with
  selective charge storage and code memory for computational imaging,''
  \emph{IEEE Transactions on Circuits and Systems I: Regular Papers}, vol.~65,
  no.~5, pp. 1555--1566, 2017.

\bibitem{sarhangnejad20195}
N.~Sarhangnejad, N.~Katic, Z.~Xia, M.~Wei, N.~Gusev, G.~Dutta, R.~Gulve,
  H.~Haim, M.~M. Garcia, D.~Stoppa \emph{et~al.}, ``Dual-tap
  pipelined-code-memory coded-exposure-pixel cmos image sensor for
  multi-exposure single-frame computational imaging,'' in \emph{2019 IEEE
  International Solid-State Circuits Conference-(ISSCC)}.\hskip 1em plus 0.5em
  minus 0.4em\relax IEEE, 2019, pp. 102--104.

\bibitem{fossum2014review}
E.~R. Fossum, D.~B. Hondongwa \emph{et~al.}, ``A review of the pinned
  photodiode for ccd and cmos image sensors,'' \emph{IEEE J. Electron Devices
  Soc}, vol.~2, no.~3, pp. 33--43, 2014.

\bibitem{mase2005wide}
M.~Mase, S.~Kawahito, M.~Sasaki, Y.~Wakamori, and M.~Furuta, ``A wide dynamic
  range cmos image sensor with multiple exposure-time signal outputs and 12-bit
  column-parallel cyclic a/d converters,'' \emph{Solid-State Circuits, IEEE
  Journal of}, vol.~40, no.~12, pp. 2787--2795, 2005.

\bibitem{park2009high}
J.-H. Park, S.~Aoyama, T.~Watanabe, K.~Isobe, and S.~Kawahito, ``A high-speed
  low-noise cmos image sensor with 13-b column-parallel single-ended cyclic
  adcs,'' \emph{IEEE Transactions on Electron Devices}, vol.~56, no.~11, pp.
  2414--2422, 2009.

\bibitem{xillybus.com}
\BIBentryALTinterwordspacing
``An fpga ip core for easy dma over pcie with windows and linux.'' [Online].
  Available: \url{http://www.xillybus.com/}
\BIBentrySTDinterwordspacing

\bibitem{oe2016}
\BIBentryALTinterwordspacing
J.~Newman, J.~Voigts, and A.~C. Lopez, ``Submillisecond latency closed-loop
  feedback with pcie prototype system,'' May 2016. [Online]. Available:
  \url{http://www.open-ephys.org/blog/2016/5/6/submillisecond-latency-closed-loop-feedback-with-pcie-prototype-system}
\BIBentrySTDinterwordspacing

\bibitem{lopes2015bonsai}
G.~Lopes, N.~Bonacchi, J.~Fraz{\~a}o, J.~P. Neto, B.~V. Atallah, S.~Soares,
  L.~Moreira, S.~Matias, P.~M. Itskov, P.~A. Correia \emph{et~al.}, ``Bonsai:
  an event-based framework for processing and controlling data streams,''
  \emph{Frontiers in neuroinformatics}, vol.~9, p.~7, 2015.

\bibitem{jonathan_newman_2019_3254431}
\BIBentryALTinterwordspacing
J.~Newman, J.~Voigts, J.~Zhang, ckemere, PhilDakin, T.~Manders, yaoyua,
  Z.~Rosen, and open ephys, ``jonnew/open-ephys-pcie: Release 1.0.0,'' Jun.
  2019. [Online]. Available: \url{https://doi.org/10.5281/zenodo.3254431}
\BIBentrySTDinterwordspacing

\bibitem{farneback2003two}
G.~Farneb{\"a}ck, ``Two-frame motion estimation based on polynomial
  expansion,'' in \emph{Scandinavian conference on Image analysis}.\hskip 1em
  plus 0.5em minus 0.4em\relax Springer, 2003, pp. 363--370.

\bibitem{lucas1981iterative}
B.~D. Lucas, T.~Kanade \emph{et~al.}, ``An iterative image registration
  technique with an application to stereo vision,'' 1981.

\bibitem{dosovitskiy2015flownet}
A.~Dosovitskiy, P.~Fischer, E.~Ilg, P.~Hausser, C.~Hazirbas, V.~Golkov, P.~Van
  Der~Smagt, D.~Cremers, and T.~Brox, ``Flownet: Learning optical flow with
  convolutional networks,'' in \emph{Proceedings of the IEEE international
  conference on computer vision}, 2015, pp. 2758--2766.

\bibitem{sun2015learning}
J.~Sun, W.~Cao, Z.~Xu, and J.~Ponce, ``Learning a convolutional neural network
  for non-uniform motion blur removal,'' in \emph{Proceedings of the IEEE
  Conference on Computer Vision and Pattern Recognition}, 2015, pp. 769--777.

\bibitem{nah2017deep}
S.~Nah, T.~Hyun~Kim, and K.~Mu~Lee, ``Deep multi-scale convolutional neural
  network for dynamic scene deblurring,'' in \emph{Proceedings of the IEEE
  Conference on Computer Vision and Pattern Recognition}, 2017, pp. 3883--3891.

\end{thebibliography}

\end{document}